\documentstyle[multicol,pra,aps,amsfonts,epsf]{revtex}

\newcommand{\cA}{{\cal A}}

\newcommand{\cF}{{\cal F}}

\newcommand{\beq}{\begin{equation}}
\newcommand{\eeq}{\end{equation}}
\newcommand{\beqy}{\begin{eqnarray}}
\newcommand{\eeqy}{\end{eqnarray}}
\newtheorem{Observation}{Observation}

\newtheorem{Definition}{Definition}
\newtheorem{Lemma}{Lemma}
\newtheorem{Theorem}{Theorem}

\newenvironment{Proof}{{\it Proof: \,}}{$\Box$ \vspace{0.3cm}}

\newenvironment{Definition*}{{\bf Definition}}{}

\makeatletter
\def\@beginTheorem#1#2{\trivlist \item[\hskip \labelsep{\bf #1\ #2}]}
\def\@opargbegintheorem#1#2#3{ \trivlist
      \item[\hskip \labelsep{\bf #1\ #2\ (#3)}]}
\makeatother

\makeatletter
\def\@beginLemma#1#2{\trivlist \item[\hskip \labelsep{\bf #1\ #2}]}
\def\@opargbeginLemma#1#2#3{ \trivlist
      \item[\hski
 Hence we have the same statements
about the increase of the supports for increasing depth, where
the local transformations are not counted for the depth.
p \labelsep{\bf #1\ #2\ (#3)}]}
\makeatother

\makeatletter
\def\@beginDefinition#1#2{\trivlist \item[\hskip \labelsep{\bf #1\ #2}]}
\def\@opargbeginDefinition#1#2#3{ \trivlist
      \item[\hskip \labelsep{\bf #1\ #2\ (#3)}]}
\makeatother

\makeatletter
\def\@beginCorollary#1#2{\trivlist \item[\hskip \labelsep{\bf #1\ #2}]}
\def\@opargbeginCorollary#1#2#3{ \trivlist
      \item[\hskip \labelsep{\bf #1\ #2\ (#3)}]}
\makeatother

\makeatletter
\def\@beginExample#1#2{\trivlist \item[\hskip \labelsep{\bf #1\ #2}]}
\def\@opargbeginExample#1#2#3{ \trivlist
      \item[\hskip \labelsep{\bf #1\ #2\ (#3)}]}
\makeatother

\def\C{{\mathbb{C}}}

\def\R{{\mathbb{R}}}

\def\N{{\mathbb{N}}}

\newcommand{\cH}{{\cal H}}

\title{Quasi-order of clocks and their synchronism\\ and quantum bounds for copying 
timing information
}

\author{Dominik Janzing\thanks{Electronic address: janzing@ira.uka.de}  and 
Thomas Beth}
\address{Institut f\"ur Algorithmen und Kognitive Systeme, Am Fasanengarten 3a,
    D--76\,131 Karlsruhe, Germany}

\begin{document}
\maketitle

\begin{abstract}  

The statistical state of any (classical or quantum) system with
non-trivial time evolution can be interpreted as the pointer of a
clock.  The quality of such a clock 
is given by the statistical distinguishability
of its states at different times. If  a clock is
used as a resource for producing  another one the latter
can  at most have  the quality of the resource. 
We show that this principle, formalized by a
quasi-order, implies constraints on  many physical
processes. Similarly, the degree to which 
two (quantum or classical) clocks are
synchronized can be formalized
by a {\it quasi-order of synchronism}.

Copying timing information is restricted by
quantum no-cloning and no-broadcasting theorems since classical clocks
can only exist in the limit of infinite energy.
We show this quantitatively by comparing the {\it Fisher timing information}
of two output systems to the input's timing information.
For classical signal processing in the quantum regime our results imply
that a signal looses its localization in time if it is amplified and 
distributed to many devices.
\end{abstract}

\begin{multicols}{2}

\section{Quantum clocks} 

Synchronizing clocks of distant parties is an important part of 
modern communication. Schemes for performing synchronization with minimal 
physical resources  have attracted attention recently 
(see e.g.\cite{GLMS} and references 
therein).
The application of these schemes 
has mostly been described 
in the context of synchronizing clocks of 
rather distant parties. Here we have in mind the miniature 
version of the scenario and recall that time synchronization takes place
on every computer chip since a common clock signal controls many 
circuits at once. In other words, since the clock signal 
triggers the execution of 
 several 
gates implemented at the same time, it realizes
essentially a clock synchronization. 

More explicitly, one can think of the following situation:
a physical signal, a light pulse or electrical pulse for instance, 
enters an amplifier.
The amplifier
produces two output signals in different directions in order to supply
two distant gates with the same clock signal.
To assure that both signals reach the corresponding gates at the same time
is an example of time synchronization.

In the common language of communication protocols, the situation
we have in mind can be
described as follows. Alice (`the source emitting the clock signal') 
sends
a physical system  that carries information about her time to Bob.
He (`the amplifier') copies the signal and sends each  copy
to another person in order to provide them with timing information.
Abstractly speaking, Alice sends a quantum system with density matrix
$\rho$ to Bob. The state $\rho$ evolves according to the system's 
Hamiltonian
$H$, i.e., it is given by
\[
\rho_t:=\exp(-iHt)\rho \exp(iHt)
\]
at the time $t$ after
Alice has prepared the state $\rho$.
Here we assume that $[\rho, H]\neq 0$, such that the states $\rho_t$ 
are not all the same. Then the actual system state $\rho_t$
contains {\it some} information about Alice's time. 
We emphasize the word `some' since it is not possible
to readout the parameter $t$, i.e., every measurement performed
on the quantum system allows only to {\it estimate} $t$.
Furthermore it is not possible for Bob to provide 
a third party (Carol) with exact copies
of $\rho_t$. Assume the clock for instance to be a two-level system
in the state  $|0\rangle + \exp(-i\omega t) |1\rangle$ with unknown $t$.
Then the problem of preparing two clocks of this type with the same
time $t$ is the problem of cloning   {\it equatorial} qubits (see 
\cite{FMWW} and references therein), 
which is only possible with a certain fidelity.
In order to be more general, consider a finite dimensional
quantum system with density matrix $\rho_t$.
To produce copies of $\rho_t$ would mean to prepare states
$\sigma_t$ on the tensor product Hilbert space  $\cH\otimes\cH$  
(if $\cH$ is the space that $\rho_t$ acts on) in such a way that
the partial traces over the first and second components
of $\sigma_t$  coincide both with $\rho_t$. This is usually referred
to as {\it quantum broadcasting}. It is possible if and only if
all the $\rho_t$ mutually commute (see \cite{Fu96} Subsection 4.3), a requirement 
that cannot be true for any non-trivial dynamics.

Can timing information be copied?  Obviously it is possible to perform
measurements on the unknown state $\rho_t$ in order to give optimal 
estimations for $t$. Then one can initialize an arbitrary large number
of  classical clocks according 
to the estimated time and distribute them to different parties. 
Here we are interested in the question whether 
the initialized classical clocks are as good as the original density matrix
$\rho_t$. In some sense, they are: if one is interested in giving
optimal estimations about $t$, each party is able to estimate
$t$ with the same error probabilities as the owner of the system $\rho_t$.
Then all parties have optimal information about $t$ 
provided that the measurement was optimal with respect to
a common optimality criterion of all parties.
However, there is another meaning of timing information
and in this broader sense not all the information of $\rho_t$ can be copied.
To illustrate this consider the equatorial qubit in the state
$|0\rangle + \exp(-i\omega t) |1\rangle$
prepared by Alice. Assume Bob receives the state,
performs  a measurement on it and initializes his clock according to
his estimated value $t'$  of $t$. Then he can prepare a new qubit in the state
$|0\rangle + \exp(-i\omega t')|1\rangle$ and hope that $t'\approx t$ 
in order to have approximatively a copy of the original state.
From Alice's point of view, Bob is not able at all to prepare 
any coherent superposition of energy eigenstates
with well-defined phase (relative to Alice's time). In this sense, he has lost
timing information in the measurement procedure.
Adopting the latter point of view timing information is the ability 
to {\it produce coherent superpositions of energy states with 
well-defined phase.}

The paper is organized as follows.
In Section \ref{Quasi} we introduce a quasi-order of clocks 
classifying resources with respect to their `worth' for
preparing coherence. As an example, we consider one mode of a light field
in a Glauber state 
as the resource clock and investigate for which states it
can provide enough timing information for producing 
a perfect coherent superposition in a two-level system:
coherent states are only suitable in the limit of infinite phonon number but
many other states in the Fock space can provide enough timing information.

In Section \ref{Accur} we illustrate how to
quantify the quality of clocks by their
 {\it Fisher timing information}.
This concept will be useful for discussing quantum bounds on copying 
timing information.
In Section \ref{Copy} we show that the Fisher timing  information 
of two copies
can only coincide with the resource's timing information
in the limit of an  infinite amount of available energy.
We derive bounds on the Fisher information of the copies
depending on the clock's energy.
This describes quantitatively the transition
from {\it quantum} timing information to {\it classical} timing information.

In Section \ref{Syn} we show that the problem of evaluating the synchronism
of two clocks can be reduced to the quality of a corresponding clock.

In this article clocks are assumed to be closed physical systems with
well-known dynamical evolution, i.e. no decoherence is considered.
Hence the clock's information  about the actual time  is only
limited by (1) inaccurate initialization of the  pointer position,
(2) quantum fluctuations of the pointer position, and (3) the periodicity
of the dynamics.

\section{Quasi-order of clocks}

\label{Quasi}

We characterize a quantum `clock' abstractly by the pair $(\rho,H)$, where
$\rho$ is the system's density matrix and $H$ its Hamiltonian.
According to the distinguishability of the time evolved states 
$\rho_t=\exp(-iHt)\rho \exp(iHt)$
different clocks have different worth. To classify the worth, we 
introduce the following
quasi-order of clocks:

The clock $(\rho,H)$ is able to prepare the clock
 $(\tilde{\rho},\tilde{H}))$, formally writing
\[
(\rho,H) \geq (\tilde{\rho},\tilde{H})
\]
if there is a completely positive trace preserving map $G$ 
such that $G(\rho)=\tilde{\rho}$ and $G$ satisfies  the covariance condition  
\[
G\circ \alpha_t =\tilde{\alpha}_t \circ G \,\, \hbox{ for all } t\in \R
\]
with the abbreviation 
\[
\alpha_t (\sigma) := \exp (-iHt) \sigma \exp (iHt)
\] 
for each density matrix $\sigma$ (and $\tilde{\alpha}_t$ defined 
similarly).

We say that two clocks are equivalent if each of both majorizes
the other in the sense of this quasi-ordering.
The equivalence class of trivial (or `worthless') clocks is 
given by all those with
$[\rho,H]=0$. They are majorized by all the other clocks.

We illustrate the intuition of this quasi-order in the following protocol:

Assume Alice sends a clock $(\rho,H)$ to Bob. Assume this system for the
moment to be a spin 3/2 system rotating about its z-axis.
Carol asks Bob to send her a clock showing as much about Alice's time as 
possible, but she complains: `please do not send the spin 
 3/2 system, I am not able to read it out.'
Since  she  can only deal with
spin 1/2 particles, she asks Alice to put as much information as possible 
about Alice's time into a spin-1/2-particle. 
All that Bob can do is mathematically  described by 
a completely positive trace preserving map $G$ 
from the set of density matrices
of spin 3/2-systems to the set of density matrices of spin-1/2-systems.
To illustrate why $G$ has to satisfy the covariance condition note that
the state of the system received by  Carol is given by
\[
\tilde{\alpha}_{t-\tau} \circ G \circ \alpha_\tau (\rho)
\]
if the time $t$ has passed since the preparation of the resource state $\rho$
and Bob was running the conversion process at time $\tau$.
Since Bob is not able to run the process at a well-defined time $\tau$, 
he applies a mixture over all $0\leq \tau \leq t$, 
whether he wants or not.
If we assume for simplicity that $t$ is large compared to the recurrence time
of the quantum dynamics this mixture can equivalently be 
described by a time covariant map $\overline{G}$.

It is interesting to note that the covariance condition also
appeared in a thermodynamic context in \cite{JWZ00}. There we introduced 
a quasi-order classifying the thermodynamic worth of energy resources.
The intuitive meaning of the covariance condition was the assumption
that it is only allowed to apply {\it energy conserving} processes
for converting one energy source into another, otherwise
one would use additional energy sources.
The reason that the covariance condition appears in both cases
is essentially the energy-time uncertainty principle:
If a process is started by a negligible amount of switching energy it
cannot be well-localized in time, i.e., the process can be started 
without using a clock.

One can easily extend the quasi-order  in such a way that it
is appropriate for classical and quantum systems.
Then one can describe each system by a unital $C^*$-algebra $\cA$ 
of observables and the states are positive functionals $\rho$ with
$\rho(1)=1$ \cite{BR1}.
Note that the expression $\rho (A)$ 
generalizes the expression $tr(\rho A)$ where $\rho$ is a density matrix.
We shall use the latter notation if we are doing usual Hilbert space quantum
mechanics 
and use the other notation if we discuss classical and quantum systems.
  The time evolution $\alpha$ on the states is given by the dual
of an automorphism group on $\cA$.
A {\it process} or {\it channel}  is described as follows. 
If the input is a state on the algebra $\cA$ and the output a state
on $\tilde{\cA}$ the process $G$ is 
the dual of a completely positive unity-preserving map
from $\tilde{\cA}$ to $\cA$.

A clock is defined
as the  pair $(\rho,\alpha)$ specifying the actual state and its 
dynamical evolution.

We set $(\rho, \alpha) \geq (\tilde{\rho}, \tilde{\alpha})$ if
there is a
process $G$
from the state space of $\cA$ to the state space of $\tilde{\cA}$ 
such that 
 $\tilde{\rho}= G(\rho)$ and $G$ is satisfying  the covariance condition
\[
G\circ \tilde{\alpha}_t =\alpha_t \circ G  \hbox{ for all } t \,.
\] 
A classical clock can for instance be given by a rotating pointer, i.e.,
a point $z$ on the 
unit circle $\Gamma\subset \C$
with the dynamical evolution 
\[
z\mapsto z \exp (i\omega t) \,.
\]
Then  the observable algebra $\cA$ is the set $C(\Gamma)$ of continuous 
functions  on $\Gamma$. The dynamical evolution of the observable 
$f$ is given by  
\[  
\alpha_t (f) (z) := f(z\exp(-i\omega t))\,.
\]
A state $\rho$ 
 on the classical algebra $C(\Gamma)$ is a 
probability measure
on $\Gamma$. If $\rho$ is the point measure of a single point
$z\in \Gamma$
the clock $(\rho, \alpha)$ has no inaccuracies at all.
In the following we will use both notations $(\rho,H)$
and $(\rho,\alpha)$ for quantum systems, whereas 
classical clocks are always denoted by $(\rho,\alpha)$.
Note that for two classical `unit circle clocks'
with the same periodicity and measures $\mu$ and $\tilde{\mu}$, respectively,
the first clock majorizes the second if and only if there is a probability
measure $\nu$ such that the convolution product $\mu \star \nu$ coincides
with $\tilde{\mu}$.

\vspace{0.5cm}

{\bf Time transfer from quantum to classical clocks}

\vspace{0.5cm}

Assume we have a quantum clock $(\rho, \alpha)$ rotating with 
the recurrence time $1$, i.e., $\rho \circ \alpha_1 =\rho$
and we want to transfer its timing information on a classical clock
with the same period. Consider for instance the classical clock
mentioned above with $\omega=2\pi$. 
Time transfer  is done by {\it measuring} the quantum system and
{\it initializing} the classical pointer to the
 position 
$\exp(-i 2 \pi t')\in \Gamma$ according to the
estimated time $t'$. The set of possible 
values $t'$ in this estimation might be
discrete or continuous and  a priori there is no restriction to the 
positive operator valued measure describing the possible  measurements.
Nevertheless one can show that each clock initialization can equivalently
be performed by a time covariant measurement.
Formally it  is described by
a family of positive operators
$(M_t)_{t\in [0,1]}$ with $\int_0^1 M_t =1$ such that
$\alpha^*_{\tau} (M_t) =M_{\tau-t}$ where the parameter $t$ is
defined modulo $1$ and
$\alpha^*$ is the dual of $\alpha$.

The fact that only covariant measurements are relevant 
can be justified by a similar argument
as above when the covariance condition for the CP-maps was motivated:
covariance describes merely the fact that the 
time of the measurement and initialization procedure is unknown.
The result that only time covariant measurements are relevant might seem 
astonishing at first sight, since 
optimal measurements of states transformed into each other by the action 
of a group are not in general group covariant \cite{PS} and optimality
has only be shown for irreducibility of the group action \cite{Da78,SBJOH}.
This shows the difference between the following tasks:
(1) Estimate the time at which the  measurement has been taken place
or  (2) initialize a clock in such a way that
one can get information about the actual time {\it later on}.

The set of covariant CP-maps $G$ from the quantum system to the
classical clock 
is in one-to-one correspondence
to the set of covariant measurements. 
This can be seen as follows.
For all set $A_t\subset \Gamma$ defined by 
$A_t:= \{ \exp(i 2 \pi t') \, | \,\, 0\leq t'\leq t\}$ we 
postulate
\[
G(\sigma) (A_t) = \int_0^t \sigma (M_{t'}) dt' 
\,\,\,\,\,\,\forall 0\leq t\leq 1
\]
for each density matrix $\sigma$.
This defines a map $G$ for each covariant POVM $(M_t)$ 
and conversely a covariant family $(M_t)$
for each covariant $G$.
This is rather intuitive since it states that the only transfer of
timing information from quantum to classical clocks is given by
measurements.
Is the worth of such 
a classical clock less than the worth of the quantum clock
that was measured? To see that in some sense, this is indeed the case
assume one provides us with the additional information
that the time $t$  is either equal to $t_1$ or to  $t_1 +1/2$.
Then the qubit is always  
able to decide which value is true, whereas generally the classical clock
can only provide probabilities for both cases.

\vspace{0.5cm}

{\bf Time transfer from classical to quantum clocks}

\vspace{0.5cm}

The set of quantum clocks that can be prepared using a given
classical clock
is easy to describe. Consider the classical  
`unit circle clock' $(\rho,\alpha)$ with period $1$ as above and 
a quantum clock
$(\tilde{\rho}, \tilde{\alpha})$ with the same period.
Assume 
\[
(\rho,\alpha)\geq (\tilde{\rho}, \tilde{\alpha})
\]
and let $G$ be the corresponding CP-map.
Consider the point measure $\delta_1$ on the point
$1\in \Gamma$. Set $\sigma:=G(\delta_1)$.
Since $G$ is time covariant and preserves convex combination
we have
\[
G(\rho):= \int_0^1 \tilde{\alpha}_t (G(\delta_1)) d\rho (t) \,.
\]
Less formally speaking every procedure for obtaining
a quantum clock from a classical one is completely defined 
by determining which quantum state $G(\delta_1)$  one wants to prepare
if the pointer of the classical clock is in position $t=0$.
Due to the covariance condition one has to prepare
the state $ \tilde{\alpha}_t (G(\delta_1))$ if the clock
is at position $t$.
This  illustrates that a classical clock can only prepare
pure  superpositions of energy states if it is supported by
discrete points.
The quantum clock $|0\rangle + \exp(-i2 \pi t) |1\rangle$ for instance
can only be prepared by the unit circle clock if $\rho$ is a single
point measure.

\vspace{0.5cm}

{\bf Time transfer from quantum to quantum clocks}

\vspace{0.5cm}

In order to illustrate that the quasi-order of clocks describes
constraints on many physical processes, we
consider a two level system initialized in its lower state $|0\rangle$.
It should be driven into a superposition state $|0\rangle +|1\rangle$
by a coherent light pulse with the appropriate frequency.
In the following we  show that the quasi-order gives bounds on the 
maximally achievable fidelity for preparing the desired pure state.
The resource clock, i.e., the optical  mode is 
the system $(|\psi\rangle\langle \psi |, H)$, where $|\psi\rangle \langle 
\psi |$ is the
coherent  state (`Glauber state') \cite{WM}
\[
|\psi \rangle := \sum_n  f_n |n\rangle
\]
with $|f_n|^2=1$ and $f_n = \exp( -|\alpha|^2/2)  \alpha^n/\sqrt{n!}$.    
The Hamiltonian $H$ acts on the photon number states $|n\rangle $ as
$H |n\rangle =n |n\rangle$ (if the frequency is assumed to be 1).
The process initializing the rotating two-level system is described by
the completely positive map $G$ from the set of density matrices 
on the Fock space to the density matrices on $\C^2$.
We write $G$ by Krauss operators 
\[
G\rho = \sum_j A_j \rho A^\dagger _j \,.
\]
with $\sum_j A^\dagger_j A_j=1$.
Due to the 
covariance condition for $G$ the operators $A_j$ can be chosen to be 
time invariant up to phase factors, 
i.e., $\exp (-i\tilde{H}t) A_j \exp (iH t)= \exp(-i\lambda t)
A_j$
where $\tilde{H}$ is the two-level Hamiltonian $H|j\rangle =j |j\rangle$ 
for $j=0,1$ and $\lambda \in \R$.
The only possibilities for $A_j$ are given by
\begin{eqnarray*}
A_0&:=& d_0 |0\rangle \langle 0| \\
A_j&:=& c_j |0\rangle \langle j-1| + d_j |1\rangle \langle j| 
\end{eqnarray*}
for $j\in \N$ and $c_j,d_j \in \C$.
The condition 
$\sum_j A_j^\dagger  A_j=1$ implies 

\begin{equation}\label{Norm} 
|c_n|^2+ |d_{n-1}|^2=1\,\,\, \forall\, n\in \N .
\end{equation}

Now we prove that the quasi-order gives 
bounds on the achievable fidelity although we will not  calculate them
explicitly.
The proof is valid for all states satisfying $f_{1} < f_0$. 

Assume that $|\psi\rangle $ allows arbitrarily perfect preparation of the 
superposition state $|\psi \rangle:=
|0\rangle + |1\rangle$. Then there is a sequence
$G_n$ of time covariant completely positive maps with 
\[
\lim_{n\to \infty}
G_n(|\psi \rangle
\langle \psi |) =|\phi \rangle \langle \phi|\,.
\]
By standard compactness arguments (each 
$G_n$ maps into a finite dimensional space)
we can assume without loss of generality 
(consider appropriate subsequences of $(G_n)$)  that 
$G(\sigma):=\lim_n G_n (\sigma)$ 
exists for all density matrices $\sigma$.
The map $G$ is completely positive and time covariant.

By assumption, we have $G(|\psi \rangle \langle\psi |)=|\phi \rangle \langle
\phi |$. Hence each $A_j |\psi\rangle $ has to be in the span of
$|\phi \rangle$.
This implies 
\begin{equation}\label{Bedi}
c_n f_{n-1} = d_n f_{n} \,\, \hbox {for all} \,\, n\in \N_0
\end{equation}
 where we
have defined $c_0:=0$. We conclude $d_0=0$. Due to eq. (\ref{Norm}) this 
implies $c_1=1$. If $f_1<f_0$ this would imply $d_1>1$ by eq. (\ref{Bedi}) 
in contradiction to 
eq. (\ref{Norm}). 

There is a rather intuitive argument showing that coherent states
with {\it large} photon numbers (i.e. large $\alpha$) are able to prepare
the equatorial qubit  almost perfectly: in the limit
of infinite photon numbers the coherent state has a classical phase
without any quantum uncertainty. One might draw the erroneous conclusion
that the statement proven above already follows 
from usual phase and photon number
uncertainty principle, since only infinite uncertainty of photon number
allows states without phase uncertainty. To see that this argument
is wrong consider the state $|n\rangle +|n+1\rangle$ which is a good
enough clock for preparing the rotating qubit. 
There are also less trivial examples. The state
\[
\frac{1}{\sqrt{4}}(|0\rangle +|1\rangle + |2\rangle +|3\rangle)
\] 
also majorizes the equatorial qubit with respect to the clock ordering\footnote{The fact that the state $\frac{1}{\sqrt{3}} (|0\rangle +|1\rangle 
+|2\rangle)$ does {\it not} majorize the equatorial qubit can be shown by 
similar arguments as above.}.
To see this define Krauss operators $A_n$ as above with
\[
\begin{array}{ccccc}
c_{2n} & = & d_{2n}& = &1 \\
c_{2n+1}& = &d_{n+1}& = & 0 
\end{array} 
\]
for $n\in \N_0$.
This illustrates that the equatorial qubit can either be prepared by
a quantum clock or by a perfect classical clock (here: the classical
phase). Note that the rotating classical 
phase $z\in\Gamma$   of the macroscopic light field 
is a physical example for the abstract classical clock discussed above.
Note that this classical  clock does only exist as a limit of quantum
systems with {\it energy} increasing to infinity.

Is it a general principle of quantum mechanics that 
{\it classical} timing  information
can only exist in the limit of infinite energy? 
In order to discuss this question consider 
a density matrix $\rho_t$ evolving according  to its Hamiltonian. 
Broadcasting of $\rho_t$ is only possible if the different
$\rho_t$ mutually commute.
Hence one can argue that timing information can only
be classical information in this example  
if one assumes (by prior knowledge about $t$)
that $t=t_1,t_2,\dots$ such that $[\rho_{t_i},\rho_{t_j}]=0$ for all
pairs $i,j$. If the distances between the possible times $t_i$ 
are decreased the required energy of the system  
goes to infinity: assume that $\rho_t$ commutes with 
$\rho_{t+\tau}$ and $\rho_t\neq \rho_{t+\tau}$.
Then the dynamical evolution has permuted some eigenvectors of $\rho_t$.
It has been shown in \cite{ML98} that a pure state $|\psi \rangle$ can 
only evolve
into an orthogonal one within the time interval $\tau$ if its mean energy
is as least $h /2\tau$.

If no prior information about the time is available 
one can use the quasi-order of clocks as a formal basis 
to derive bounds on the timing information of two clocks
obtained from one resource clock.
The task is to find
a clock of the form
\[
(\tilde{\rho}, \tilde{H}_1\otimes 1 + 1\otimes \tilde{H}_2)
\]
that can be produced from the original clock $(\rho,H)$ with the 
following property:
The subsystems 
$(\tilde{\rho}_1,\tilde{H}_1)$ and $(\tilde{\rho_2},\tilde{H}_2)$
obtained by tracing out the right or left system, respectively, 
are {\it good clocks}.  
In contrast to usual cloning problems, the task here is not to obtain optimal
copies of $\rho$ but to put as much timing information as possible
into both systems. The states $\tilde{\rho}_1$ and $\tilde{\rho}_2$ 
can differ considerably from $\rho$.
In order to investigate the maximal timing information in both
components we have to quantify the information.
There are many possibilities for doing so, the quantity we chose in the next
section has many useful properties that justify
to consider it as a resource for physical processes.

\section{Fisher timing information}

\label{Accur}

In order to quantify the accuracy of a clock consider
the following situation in every-day life: 
imagine
a classical clock having such a wide pointer that one is not able
to read out its time exactly.
Here it is straightforward to define the {\it time error} as the pointer's
width $w$ divided by its velocity $v$.
To translate this to the quantum clock
$(\rho,H)$  we consider an observable represented by the POVM 
$M:=(M_\lambda)_{\lambda \in \R}$ defined in such a way that
$tr ( \rho M_\lambda)$ is the probability for measuring 
values less or equal to $\lambda$ (Note that
this definition does not coincide with the definition in Section \ref{Quasi}
where $\rho(M_t)$ was the probability density for measuring $t$.
Here we do not restrict our attention to covariant measurements and 
cannot assume that a probability density describing the measurement exists.).
The
expectation value of the measurement outcomes is 
$tr( \rho \int \lambda \, d M_\lambda)$ and  moves with the velocity
\[
v:=tr(-i [H,\rho] A)\,
\]
with $A:= \int \lambda \,1\, d M_\lambda$.
The measurement values have a 
distribution with standard deviation
\[
w:=\sqrt{ \int tr(\rho (\lambda \,1\, - (tr(\rho A))^2 \, dM_\lambda)} \,.
\]
We can consider the quotient $v/w$  of the
standard deviation and the time derivative of the expectation value
as an analogue of the time error above.
The problem of finding 
a POVM minimizing this error and finding bounds
on the error is a quantum estimation problem that has already been studied
in the literature \cite{BC94,BCM96,Hole,Hel}. We rephrase some of the results
in our setting and relate them to our quasi-order.
As already noted in \cite{BC94}
the optimal  POVM  can always be chosen to be projector-valued.
We give a different proof
by showing that 
the projector valued measure defined by the spectral resolution of
$A$ can never be worse than the POVM $M$.
This can be seen as follows:
The expectation values and their time derivatives 
of both measurements coincide.
The variances differ by $tr(\rho (\int \lambda^2 dM_\lambda)-A^2)$.
Some calculations show
\begin{eqnarray}\label{integral}
&&\int \lambda^2 \,dM_\lambda-A^2 = \nonumber \\
&&\int \Large(\lambda \,1\, -\int \mu \,1\,\,d M_\mu)\,\, d M_\lambda \,\,
(\lambda \,1\, -\int \mu \, 1\, \, dM_\mu )\Large\,.
\end{eqnarray}
This notation may require some explanations. 
Assume the POVM consists
of a finite set $Q_j$  of positive operators with $\sum_j Q_j=1$
corresponding to the measurement outcomes $\lambda_j$. Then
the corresponding expression is
\[
\sum_j \Large((\lambda_j -\sum_i \lambda_i Q_i ) Q_j 
 (\lambda_j -\sum_i \lambda_i Q_i )\Large)\,.
\]
In the case of an infinite number of results,
$dM_\lambda$ is the analogue of $Q_j$ and the summation is substituted 
by an integral. 

The right hand side of eq. (\ref{integral}) is a positive operator
since an operator of the form $B C B $ with self-adjoint operators $B$ 
and $C$ is always positive  and an integral of positive operators is
positive too.

We define the Fisher timing information of a {\it classical or quantum clock} 
as the inverse of the squared time error for the optimal measurement.
Formally, this reads as follows.

\begin{Definition}
Let $(\rho,\alpha)$ be a clock, i.e., $\rho$ is a state on a
$C^*$-algebra $\cA$ of observables and $\alpha:=(\alpha_t)_{t\in \R}$ 
be the group defining the time evolution of states.
Then its {\bf Fisher timing information}  $F$ is defined as
\[
F(\rho,\alpha):= \sup_{A} \frac{d}{dt} \alpha_t (\rho) (A)|_{t=0})^2 /\rho(A^2)   
\]
where one optimizes over all self-adjoint operators $A\in \cA$.
\end{Definition}

To justify this terminology  we show that $F$ reduces to the usual Fisher 
information when applied to  classical clocks.
Let the state be a probability distribution on a phase space $X$. 
Assume that it is given by a density function 
$x\mapsto p(x)$.  At time $t$ we have the density function 
$x \mapsto p_t(x)$. Let $x\mapsto f(x)$ be an observable, i.e., a continuous
 real function on the phase space. The time 
derivative of its expectation value  is given by
\[
\int f(x) \frac{d}{dt} p_t (x) dx
\] 
and its variance by
\[
\int f^2(x) p_t(x) dx \,,
\]
if the expectation value of $f$ 
is assumed to be zero without loss of generality.
For technical reasons we assume $p_t(x)\neq 0$ for all $x$.
Then we can define an inner product on the set of real  continuous  
bounded functions on $X$ by 
\[
\langle f|g\rangle :=\int f(x) g(x) p(x) dx \,.
\]
We have to maximize the expression
\[
\frac{(\langle f | \dot{p}_t/p_t \rangle)^2}{ \langle f|f\rangle }
\] 
with $\dot{p}_t=d/dt p_t$.
For geometric reasons the function
\[
f:=\frac{\dot{p}_t}{p_t} = \frac{d}{dt} (\ln p_t)
\]
is optimal. Then $F$ is given by the usual Fisher information
\[
\int \frac{(\dot{p}_t (x))^2 }{p_t(x)} dx \,,
\]
a quantity that is well-known
in classical statistics. It is
decisive as a  bound on the error of estimators \cite{Cra}, the so-called
{\it Cram\'{e}r-Rao bound}.

For finite dimensional quantum systems the optimization
can be performed similarly:
In order to maximize the expression
$tr (\dot{\rho} A)/ tr( \rho A^2) $ we can introduce a bilinear form 
\[
\langle A |B\rangle := tr( A \Delta_\rho B)
\]
on the set of self-adjoint operators introducing the super operator
$\Delta_\rho$ with
$\Delta_\rho B:=(\rho B +B \rho) /2$. For technical reasons we assume
$\rho$ to have full rank such that $\Delta_\rho$ has an inverse.
Following \cite{Hel,Hole} we obtain the `symmetric logarithmic 
derivative' $\Delta^{-1}_\rho \dot{\rho}$ as optimal observable
with $\dot{\rho}:= i[\rho,H]$. We conclude the following.

\begin{Theorem}
The Fisher timing  information of a quantum clock $(\rho,H)$ 
on a  Hilbert space
is given by
\[
F(\rho,H)= tr( \dot{\rho} \Delta_\rho^{-1} \dot{\rho} ) 
\]
with $\Delta^{-1}_\rho$ and $\dot{\rho}$ as above. 
\end{Theorem}

As noted in \cite{BC94} this expression appears also 
if one calculates the Bures distance between the states $\rho$ and
$\rho_{dt}$ for infinitesimal time $dt$:
\[
d_{Bures} (\rho, \rho_{dt}) =  tr( \dot{\rho} \Delta_\rho^{-1} \dot{\rho} ) 
(dt)^2 \,.
\]
Introducing the fidelity \cite{Fu96}
\[
f(\rho_1,\rho_2):=\inf_{M} \sum_b \sqrt{\rho_1 (M_b)} \sqrt{\rho_2(M_b)}\,, 
\]
where $M:=(M_b)_b$ is an arbitrary POVM,  
one can write the Bure distance as \cite{Fu96}
\[
d^2_{Bures}(\rho_1,\rho_2) = 2-2 f(\rho_1,\rho_2) \,.
\]
Since the fidelity is clearly monotone with respect to
completely positive maps the Bure distance is monotone too.
We conclude that Fisher timing information is
monotone with respect to our quasi-order of clocks, since
the state $G(\rho)$ evolves to the state $G(\rho_{dt})$ 
for every time covariant CP-map $G$.    
Formally we state the following:

\begin{Lemma}
For each two quantum or classical clocks  we have:
\[
(\rho,\alpha) \geq (\tilde{\rho},\tilde{\alpha})
\]
implies
\[
F(\rho,\alpha)\geq F(\tilde{\rho}, \tilde{\alpha})\,.
\]
\end{Lemma}

Similar arguments show that Fisher timing
information is constant according to the system's Hamiltonian evolution, i.e.,
$F(\rho,\alpha)=F(\alpha_t(\rho), \alpha)$.
This justifies to consider it as a convenient measure
characterizing the quality   of {\it resources} for preparing {\it 
time-localized
signals}. This is even more justified by noting that
the quantity 
is additive with respect to 
the composition of independent quantum clocks:

\begin{Lemma}\label{Addi}
Fisher information of quantum systems is an additive quantity, i.e.,
\[
F(\rho\otimes \tilde{\rho} , H\otimes 1 +1 \otimes \tilde{H})=
F(\rho,H) + F(\tilde{\rho},\tilde{H})
\]
\end{Lemma}

\begin{Proof}
As already noted in \cite{BCM96} the optimal observable
of the composite system is given by the sum 
\[
\Delta_\rho \dot{\rho} \otimes 1+ 1\otimes \Delta_{\tilde{\rho}} 
\dot{\tilde{\rho}} \,.
\] 
Then the statement follows immediately.
\end{Proof}

Similarly,
for classical clocks with a probability distributions $p$ and
$\tilde{p}$  moving 
on the phase spaces $X$ and $\tilde{X}$  
the optimal observable $(x,y)\mapsto f(x,y)$ 
on $X\times \tilde{X}$ is
given by the sum
\[
f(x,y) =\frac{\dot{p}_t(x)}{p_t(x)} + \frac{\dot{\tilde{p}}(y)_t}{\tilde{p}
(y)_t} \,.
\]
Here the additivity 
of Fisher timing information
follows straightforwardly.
We conjecture that additivity is even true for the composition 
holds even for the composition of quantum with classical 
clocks.

\vspace{0.5cm}

{\bf Examples}

\vspace{0.3cm}

We give two examples where Fisher timing information can be calculated
explicitly.

\begin{enumerate}

\item
Let $p_t$ be a Gauss distribution with standard deviation $\sigma$ 
moving with the velocity $v$ 
on the real line, i.e.,
\[
p_t(x):=\frac{1}{\sqrt{2\pi} \sigma} \exp\Large(- \frac{(x -tv)^2}{2\sigma^2}
\Large)  \,.
\]
Then $F$ is given by
\begin{eqnarray*}
F&=&\int \Large(\frac{d}{dt}p_t(x) \Large)^2 \frac{1}{p_t(x)}dx \\
&=& \int \Large( \frac{d}{dt} \ln  p_t(x) \Large)^2 p_t(x) dx =
\frac{v^2}{\sigma^2}\,. 
\end{eqnarray*}

\item
Let $\rho:=|\psi\rangle \langle \psi |$ be a pure state  
of a quantum system with Hamiltonian $H$. Then
the Fisher timing information is given by  $4$ times the
 variance of the
energy values \cite{BCM96}, i.e.,
\[
F=4(\langle \psi | H^2|\psi \rangle -(\langle \psi |H|\psi \rangle)^2)\,.
\]

\end{enumerate}

Note that the intuitive meaning of the quantity $F$ becomes even more evident
if one considers a large number $n$ of copies of the same clock:
\begin{eqnarray*}
(\rho,H)^n:&=&
(\rho\otimes \dots \otimes \rho, H\otimes 1 \otimes \dots \otimes 1 \\ &+&
1\otimes H\otimes 1\otimes \dots \otimes 1 
+\dots +1\otimes \dots \otimes 1 \otimes H)\,.
\end{eqnarray*}
First note that the error of the large clock is decreasing
in the order of $1/\sqrt{n}$, i.e., the error probability for distinguishing
between $\rho^{\otimes n} $ and $\rho^{\otimes n}_{t/\sqrt{n}}$ does 
neither converge to $0$ nor to $1$ (for details see \cite{BCM96,BC94}).

Then consider observables of the type
\begin{eqnarray}\label{Fluk}
A_n&:=&\frac{1}{\sqrt{n}} (A\otimes 1\otimes \dots \otimes 1\\&+& 
1\otimes A \otimes 1 \otimes  \dots \otimes 1 + \dots +1\otimes \dots \otimes 1\otimes A)\,, \nonumber
\end{eqnarray}
where 
$A$ is any self-adjoint operator acting on the single clock's Hilbert space
with $\rho(A)=0$.
Due to the central limit theorem, $A_n$ is approximatively Gauss distributed
for large $n$. The standard deviation of $A_n$ is given by $\sqrt{\rho 
(A^2)}$.
Easy calculation shows that
\[
\lim_{n\to \infty} (\rho_{t/\sqrt{n}})^{\otimes n} 
(A_n)=
t \,\dot{\rho} (A)\,.
\]
This has a very intuitive implication:
on the time scale $1/\sqrt{n}$ the expectation value of 
the Gauss distribution  moves
with velocity $\dot{\rho} (A)$. If one allows only measurements of the type
$A_n$  the only criterion for the distinguishability
of the states
$\rho_{t/\sqrt{n}}$ (for different values of $t$) is the quotient of the 
velocity and  $\dot{\rho} (A)$ the  $\sqrt{\rho (A^2)}$, i.e., the
square root of the Fisher timing information.

Remarkably, observables of the type $A_n$ can be considered as {\it 
approximate} time
operators in the following sense.  
Multiply the total Hamiltonian of the system by the factor $1/\sqrt{n}$ and
note that this rescaled Hamiltonian $H_n$ is also an observable of
the type  eq. (\ref{Fluk}). In a suitable sense, the commutator
\[
[A_n, H_n]=\frac{1}{n}[A,H] \,.
\]
converges to $\rho ([A,H])$ times the identity operator in the 
limit $n\to \infty$. This is made precise in so-called 
{\it non-commutative central limit theorems} \cite{GVV89,GV89}. 
In this approach the
distribution of the observables $A_n$ is given by quasi-free states on a 
CCR-algebra (`algebra of fluctuations'). 
In the  CCR-algebra the corresponding limit observables
$A_\infty$ and $H_\infty$ satisfy 
\[
\exp(iH_\infty t) A_\infty \exp( -iH_\infty t) = A_\infty + c t1
\]
with $c:= \rho (i[H,A])$. 
 
This indicates the quantum estimation problem of distinguishing 
the states $\rho^{\otimes n}_{t/\sqrt{n}}$ for different $t$
becomes more and more related to the estimation problem in a quantum Gaussian
channel (for Gaussian channels see \cite{Ho73}).

\section{Copying time-localized signals}

\label{Copy}

Since Fisher timing information seems to be a convenient quantity
to measure the quality of a clock (at least one aspect of
it) it is straightforward to ask whether it is possible to
produce two clocks from one resource clock such that
the timing information of each is the same as the resource's 
timing information.
Formally speaking we have a resource clock
$
(\rho,H)
$,
and look for a clock 
\[
(\sigma, H_1\otimes 1 +1\otimes H_2)
\]
acting on a joint Hilbert space such that the clocks 
$(\rho_1,H_1)$ and $(\rho_2,H_2)$ given by
the partial traces over the Hilbert space components 
have a large amount of Fisher information and
\[
(\rho,H) \geq (\sigma, H_1\otimes 1 +1\otimes H_2)\,. 
\]
Note that the states of the produced 
clocks are allowed to be highly correlated.
Hence one cannot argue that the Fisher information of the resource 
has to be divided
between both clocks, i.e, we may have
\[
F(\rho_1,H_1)+F(\rho_2,H_2) > F(\rho,H) \,.
\]
Here we address the question whether it is possible that
both clocks are of the same quality as the resource, i.e.,
whether one can have
\begin{equation}\label{Frage}
F(\rho_1,H_1)=F(\rho_2,H_2)=F(\rho,H)\,.
\end{equation}

The answer is that it is never possible if the energy of the produced
clocks is bounded.
To see this
assume that $H_1$ and $H_2$ are positive operators.
Let $T_1$ and $T_2$ be the optimal time observables for
$C_1:=(\rho_1,H_1)$ and $C_2:=(\rho_2,H_2)$, respectively. 
We assume $\dot{\rho}_1 (T_1) = \dot{\rho}_2(T_2)=1$ and 
$\rho_1(T_1)=\rho_2 (T_2)=0$ without loss of generality.
Then the values of the  
Fisher timing information $F_1$ and $F_2$ of $C_1$ and $C_2$
are given by $1/\rho_1 (T_1^2)$ and $1/\rho_2(T_2^2)$, respectively.
One has
\begin{eqnarray*}
&&\rho_1 (T_1^2) + \rho_2(T_2^2)\\ &=& \frac{1}{2} \Large(
\sigma ((T_1\otimes 1+1 \otimes T_2)^2) +
\sigma ((T_1\otimes 1- 1\otimes T_2)^2)\Large).
\end{eqnarray*}
Note that $\dot{\sigma}(T_1\otimes 1+1\otimes T_2)=2$. Therefore
the variance of this observable can not be less than $4$ divided by the
Fisher information $\tilde{F}$ of the state $\sigma $. Due to the monotony
of Fisher information with respect to the quasi-order we have  
$\tilde{F} \leq F$, where $F$ is the Fisher information of the resource clock.
Hence we find
\begin{eqnarray}\label{entsch}
1/F_1 +1/F_2 &\geq & 2/F + \sigma ((T_1\otimes 1-1\otimes T_2)^2)\\
&=& \nonumber 2/F + V_{\sigma} ( T_1\otimes 1-1\otimes T_2)\,,
\end{eqnarray}
where  we have introduced the notation 
$V_\sigma(A)$ for the variance of any observable $A$ in the state $\sigma$, i.e.,
\[
V_\sigma(A):= \sigma (A^2) -(\sigma (A))^2 \,.
\] 
By Heisenberg's uncertainty relation we have
\begin{eqnarray*}
&&V_{\sigma} (T_1\otimes 1 -1\otimes T_2) \,\,
V_{\sigma}(H_1\otimes 1-1\otimes H_2)\\  &\geq &
\frac{1}{4} \Large(\sigma ([T_1\otimes 1 -1 \otimes T_2 , 
H_1\otimes 1-1\otimes H_2])\Large)^2
\\&=&
\frac{1}{4}\Large(\dot{\sigma} (T_1\otimes 1 + 1\otimes T_2)\Large)^2 
=1 \,.
\end{eqnarray*} 
We have
\begin{eqnarray*}
&&V_\sigma (H_1\otimes 1- 1\otimes H_2) \\
&\leq& \sigma ((H_1\otimes 1- 1\otimes H_2)^2) \leq 
\sigma ((H_1\otimes 1 +1\otimes H_2)^2)\,,
\end{eqnarray*}
where the last inequality follows since $H_1$ and $H_2$ are positive 
operators by assumption.

We conclude
\begin{eqnarray*}
&&\sigma ( (T_1\otimes 1 - 1\otimes T_2)^2) \\&\geq& 4/(\sigma ((H_1\otimes 1 
+1\otimes H_2)^2 - (\sigma (H_1\otimes 1 -1\otimes H_1))^2).
\end{eqnarray*}
With eq. (\ref{entsch}) we have
\[
1/F_1 + 1/F_2 \geq 2/F + 2/\sigma ((H_1\otimes 1 +1 \otimes H_2)^2)\,.
\] 
We introduce the abbreviation
\[
\langle E^{2} \rangle := \rho ((H_1\otimes 1 +1 \otimes H_2)^2)
\]
for the expectation value of the square of the output's energy.
Then we can  write 
\[
1/F_1 +1/F_2 \geq 2/F + 2/\langle E^{2} \rangle\,.
\]
Assume one wants to have the same  Fisher timing information
in both outputs, i.e.,
$F_1=F_2$.
Then one obtains
\begin{equation}\label{boundend}
1/F_1 \geq 1/F +1/\langle E^{2} \rangle \,.
\end{equation}
This implies that the timing information of the target systems 
is less than the information of the resource by an amount 
determined by the square of the available energy.
One can see that perfect copy of the timing information, i.e.,  
$F_1=F_2=F$ is only possible in the limit $\langle E^{2}\rangle \rightarrow \infty$.
In particular if the energy of the system is bounded by
a maximal energy value $E_{\max}$ the loss of timing information
in the copies can be expressed in terms of $E_{\max}$.
Consider for example the case that the output systems are 
coherent light pulses with  frequency $\nu$.
Its energy is Poisson distributed. The expectation value of its energy
is $Nhv$ if $N$ is the average photon number. The variance is
$Nhv$ as well. Hence we find 
\[
\langle  E^2 \rangle =Nhv^2 +Nhv
\]
for the expectation value of the energy  squared. 
Then equation (\ref{boundend}) 
gives a lower bound on the loss of timing information
of the copies compared to the resource (depending on the mean 
photon number).

Assume now that the joint state $\tilde{\rho}$ of the target systems
is pure. Then the covariance of the energy values   coincides with 
the Fisher information  
of the joint output state.
We can write
\[
\sigma ((H_1\otimes 1+ 1\otimes H_2)^2)= 
\langle E\rangle ^2 +F(\sigma ,H_1\otimes 1 +1\otimes H_2)\,,
\]
with the abbreviation 
\[
\langle E\rangle:= \tilde{\rho} (H_1\otimes 1+ 1\otimes H_2) \,.
\]
Since the Fisher information of the joint output state 
cannot be larger than $F$ we obtain
\[
1/F_1 \geq 1/F +1/(F+\langle E \rangle ^2).
\]
The copies can only obtain the full timing information of the input
in the limit of infinite average energy (see Fig. 1). 
Since $\langle E^2 \rangle $ can 
be infinite even for finite  $\langle E\rangle$ the bound for pure states 
is considerably
tighter than for mixed states.

\begin{figure}
\centerline{
\epsfbox[0 0 232 135]{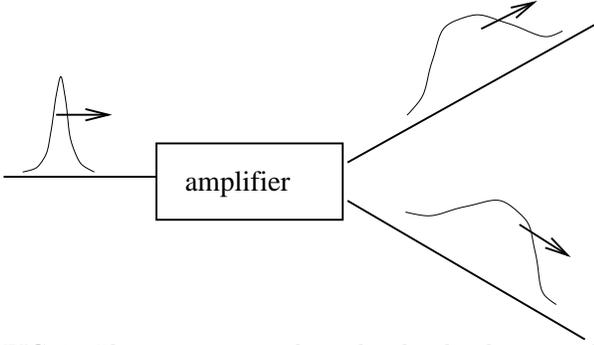}
}
\caption{The outgoing signals are less localized in time then the input.
The functions can be interpreted as the probability distribution of the 
signal's time of arrival at a certain point. The clock consisting
of  both outputs
cannot be better than the clock defined by the input (with respect 
to the quasi-order).}
\end{figure}

Note that the assumption of limited energy is indeed necessary.
In the following we sketch an example of an input signal
of large energy that produces two outgoing signals 
each of it having almost the same Fisher timing information as the
input.
Let $\cH:=L^2(\R^+)$ be the square-integrable functions on the set
of non-negative numbers. Define an operator 
$A : \cH \rightarrow \cH\otimes \cH \equiv L^2(\R^+\times \R^+)$ by 
\[ 
(A \psi) (x,y):= \left\{\begin{array}{ccc} 
\psi (x+y) /\sqrt{x+y}  & \hbox{ for  }  & x,y \geq 0 \\
0 &  \hbox{ else } & \end{array}\right.
\]
Since $A$ is an isometry we have $A^\dagger A=1$, hence the map
\[
G(.):= A (.) A^\dagger
\]
is completely positive and trace preserving.  
For both copies of $\cH$ assume the Hamiltonian
to be given by the multiplication operator 
\[
(H\psi) (x):= x \psi (x) \,.
\] 
Then $G$ is clearly time covariant, since $A$ maps states onto states with the
same energy distribution.
The operator $A$ maps the wave function of the in-going signal 
on the joint wavefuction of the outgoing signals.

Assume now the in-going wave-function 
$|\psi (x)|^2$ to be a Gauss distribution with expectation value
$E$ and standard deviation $\Delta E$ such that $E>>\Delta E$.
Its Fisher timing information is given by $F=4(\Delta E)^2$ due to Example 2
in Section \ref{Accur}.
The wave package $(x,y) \mapsto A\psi (x,y)$ is 
essentially
localized on a strip of length $\sqrt{2}E$ and width in the order of 
$\Delta E$.
Along the lines $x=y+c$ with $|c| << E$  
the function $(x,y) \mapsto |A\psi (x,y)|^2$ is approximatively a
Gaussian wave package
with standard deviation $\Delta E/\sqrt{2}$.
Introducing the isometric Fourier transformation
\[
\cF: L^2(\R^+) \rightarrow L^2(\R)
\]
with
\[
\cF (f) (k):= \frac{1}{\sqrt{2\pi}} \int f(x) \exp(-ikx) dx
\]  
the probability distribution according to
\[
|(\cF \otimes \cF )A\psi (x,y)|^2
\]
is  for large $E$ essentially localized on  the line $x=y$.
Along this line it is a Gauss distribution. To calculate its
standard deviation note that the spread of
of $|A\psi|^2(x,y)$ 
along
the $x=y$-line is $\Delta E/\sqrt{2}$. Hence we obtain the value 
$\sqrt{2}/(2 \Delta E)=1/(\sqrt{2} \Delta E)$.

To estimate the timing information of each of the signals we
define an operator $\hat{T}$ on  $L^2(\R)$ by
\[
\hat{T}\psi (x):= x\psi (x)
\]
and an operator $T$ on $L^2(R^+)$ by
\[
T:= \cF^\dagger \hat{T} \cF \,.
\]
Now we will show that  $T\otimes 1$ is an
operator with the property that its variance is approximatively
equal to $1/(4(\Delta E)^2)$ and its expectation 
values has time derivative $1$.
In order to avoid lengthy technicalities,
we sketch this by hand-waving arguments.

To see that the time derivative of the expectation values of
$T\otimes 1$ is  equal to $1$ note that the
dynamical evolution on each Hilbert space component $L^2(\R)$ 
is simply a translation by $t$ after the time $t$.
Hence the time derivative  is $1$  for {\it every} state.
There is, however, a difficulty in calculating the variance of
the observable $T\otimes 1$ with respect to  the state $A\psi$.
One would like to make use of the fourier transformation
and calculate the variance of $T$ by calculating the variance 
of $\hat{T}$ with respect to the fourier transformed wave-package.
Unfortunately this is wrong 
because $\cF$ is not unitary and we have
\[
T^2=\cF^\dagger 
\hat{T} \cF\cF^\dagger T \cF^\dagger \neq \cF^\dagger \hat{T}^2 \cF \,.
\]
Note that the operator $\cF\cF^\dagger$ is not the identity but the
projector onto the image of $\cF$.
However, this difference becomes irrelevant in the limit of
$E\to\infty$. The wave-function $(\cF \otimes \cF) A\psi$ contains 
essentially large positive frequencies since $E$ is large.
Hence $(\hat{T} \otimes 1) (\cF\otimes \cF) A\psi$ contains essentially
positive frequencies.
We conclude that the variance of 
\[
\hat{T}\otimes 1
\]
with respect to the fourier
transformed wave-function $(\cF \otimes \cF) A\psi$
is a {\it good approximation} for the variance of 
\[
T\otimes 1
\]
and the error 
goes to zero for $E\to\infty$.
The variance of $\hat{T}\otimes 1$ is easy to see in the limit 
of $E\to\infty$.
Note that is is given by the variance of the $x$-values according
to the probability density
$|(\cF \otimes \cF)A\psi |^2 (x,y)$ which is 
more and more localized on the $x=y$-line for large $E$. As we have 
already argued, the statistical distribution along this line is Gaussian
with standard deviation $1/(\sqrt{2} \Delta E)$. The standard deviation of
the $x$-values is therefore given by $1/(2 \Delta E)$ and the corresponding
variance by
$1/(4 (\Delta E)^2)$.
The Fisher timing information of each signal is therefore approximatively
given by $4 (\Delta E)^2$, the same value as we have assumed for 
the in-going  signal.

Note that this does not mean that the original state $|\psi\rangle$ 
would have been cloned approximatively. The joint state $A\psi$ is highly entangled as the measurement values of the time observables
$T\otimes 1$ and $1\otimes T$ are almost perfectly correlated.
 Hence the reduced states for both systems
are highly mixed. This shows that {\it Fisher timing information} 
can approximatively 
be copied although the {\it coherence} has mostly been destroyed.

\section{Clock synchronism and relative timing information}

\label{Syn}
So far we have only discussed timing information with respect to an
`absolute' clock that is not considered explicitly.
In most of the applications the absolute time is less relevant
than the fact that two parties have the same time, i.e., 
their clocks are 
synchronized. It is obvious that the problems of clock synchronization and
measuring the absolute time are strongly related.
Assume for instance that Alice has two clocks, one  showing the absolute time 
without any error and the other showing the time approximatively.
She sends the inaccurate clock to Bob.
Then the clocks of Alice and Bob are more or less synchronized
and the quality of the synchronism is directly given by the quality of
the transfered clock. Conversely, assume that Alice and Bob
have a clock each showing approximatively the same time
but the relation between the time of these clocks to the absolute time is
totally unclear. Now we tell Alice the absolute time.
She compares this value to the time on her clock and tells
Bob the difference. Bob's information about the absolute time 
after this protocol is directly given by the quality of the synchronism
of both clocks. In this classical setting, the problems of
transferring information about the absolute time and
synchronizing their relative time are exactly equivalent.
Alice can use 
the correlations between both clocks for sending timing information
to Bob by sending usual (non time-like) information to Bob.
The transfered information is no timing information
since it does not change during  its journey through the space.

In the following we will restrict our attention to periodic clocks
with a common period length $1$. This assures that invariant
states of the joint systems can always be obtained by averaging over the
states
of a full period. By this we avoid lengthy 
discussions on ergodic theory
concerning the existence of time means and related questions. 
With this restriction, the classical situation can be illustrated by 
two unit circle clocks rotating with the same frequency $\omega=2\pi$.

At any moment, the position  $z_A$ of Alice's pointer is 
randomly distributed and the position $z_B$ of  Bob's pointer too.
We assume equal distribution on $\Gamma$ for
the probability distribution of $z_A$ and for $z_B$ if they are 
considered separately. This expresses the fact that the clocks contain
no information about the absolute time. The fact that they are more or less
synchronized can be expressed by a joint probability measure on
$\Gamma\times \Gamma$ 
such that there is a correlation between $z_A$ and $z_B$.
We assume that the conditional probability for measuring
$z_A$ at Alice's clock provided that $z_B$ is Bob's time  is given by
\[
p(z_A | z_B)=q(z_Az_B^{-1})
\]
with a probability density $q:\Gamma \rightarrow \R^+$.
Due to the time  invariance of the joint state
we have $p(z_B |z_A)=p(z_A|z_B)$.
If the clocks are well synchronized the probability density $q$
is essentially  supported around a specific value $z\in \Gamma$
expressing the fact that the difference between Alice's and Bob's clock
is approximatively known to both.
Now we tell Alice the absolute time $t_0$. She will use this information
to calculate the difference between the true time
and her clock. In our formal setting it is given by the value
$z_A \exp(-i\omega t_0)$. Alice sends this value to Bob. He
can turn his pointer around according to this difference value.
The position of his pointer will now be distributed according to the 
probability density $q$.

For quantum systems the situation seems to be more
difficult at first sight: Alice's and Bob's clocks
may be classically correlated or entangled. If the synchronization
is given by quantum correlations, it is not clear to what extent
synchronism is destroyed by measurements on Alice's or Bob's
clock. Up to now, it is not even clear how to define synchronism
at all.
Here we suggest a formal definition of synchronism that
allows to compare the quality of synchronization of different bipartite
systems in the same way as we have compared clocks.

\begin{Definition}
A synchronism is a joint state $\rho$ on a bipartite system
$A\times B$ with separate time evolutions $\alpha$ and $\beta$ on
$A$ and $B$ such that $\rho$ is invariant with respect to 
$\alpha\otimes\beta$. Formally, we will denote a synchronism
by $(\rho,\alpha,\beta)$.
\end{Definition} 

Synchronism can be compared as follows.

\begin{Definition}
The bipartite system $A\times B$ is said to be better than 
or equally synchronized as the system $\tilde{A} \times\tilde{B}$, formally
denoted by
\[
(\rho,\alpha,\beta) \geq (\tilde{\rho},\tilde{\alpha},\tilde{\beta})
\] 
if there is a process $G$ converting $\rho$ to $\tilde{\rho}$ 
satisfying the covariance condition
\[
G \circ(\alpha^{-1}\otimes \beta) =(\tilde{\alpha}^{-1} \otimes \tilde{\beta})
\circ G \,.
\]
\end{Definition}

The intuitive meaning of the covariance condition is that the communication
protocol implementing the joint transformation $G$ on the bipartite system
is invariant with respect to a relative time displacement between Alice's 
actions and Bob's actions, i.e., Alice and Bob do not require external
synchronized clocks in order to implement $G$.
The reason for restricting the attention to stationary states $\rho$ is 
is that a non-trivial time evolution of the joint state would mean
that the bipartite system carries some information about the {\it absolute} 
time.
This would be confusing in the following considerations.
It is natural to define an equivalence relation on synchronizations by
\[
(\rho,\alpha,\beta) \equiv (\tilde{\rho},\tilde{\alpha},\tilde{\beta}) 
\]
if 
\[
(\rho,\alpha,\beta) \geq (\tilde{\rho},\tilde{\alpha},\tilde{\beta})
\]
and
\[
(\tilde{\rho},\tilde{\alpha},\tilde{\beta}) \geq (\rho,\alpha,\beta) \,.
\]

We observe:

\begin{Observation}
For two synchronizations we have
\[
(\rho,\alpha,\beta) \geq (\tilde{\rho},\tilde{\alpha},\tilde{\beta})
\]
if and only if 
\[
(\rho,\alpha^{-1} \otimes \beta) \geq 
(\tilde{\rho},\tilde{\alpha}^{-1}\otimes \tilde{\beta})
\]
with respect to the quasi-order of clocks.
\end{Observation}

This is trivial from the mathematical point of view.
Note that
\[
(\rho,\alpha^{-1} \otimes \beta) \geq 
(\tilde{\rho},\tilde{\alpha}^{-1}\otimes \tilde{\beta})
\]
if and only if 
\[
(\rho,\alpha^{-1/2} \otimes \beta^{1/2}) \geq 
(\tilde{\rho},\tilde{\alpha}^{-1/2}\otimes \tilde{\beta}^{1/2})\,,
\]
since these clocks are moving with half of the velocity
than those above.
Observe furthermore that we have
\[
(\rho,\alpha^{-1/2} \otimes \beta^{1/2}) \equiv (\rho, 1\otimes \beta)
\]
since we can define a process  $G$ by
\[
G:=\int_0^1 \alpha_t \otimes \beta_t dt \,,
\]
satisfying $G(\rho)=\rho$ and $G \circ (\alpha^{-1/2} \otimes \beta^{1/2})
=(1\otimes \beta) \circ G$ and 
$(\alpha^{-1/2} \otimes \beta^{1/2}) \circ G=G \circ (1\otimes \beta)$. 

We would like to given an intuitive meaning to some of those clocks.
For doing so, we show in the following that 
the clock $(\rho,1\otimes \beta)$ is the best clock
that Bob can obtain if we tell Alice the absolute time 
and allow Alice and Bob to communicate 
by a classical {\it and} quantum channel without transferring clocks.
This point has to be clarified. If Alice would sent a rotating two-level 
system to Bob this would be a {\it clock transfer}. 
But we {\it do allow} Alice to send
a {\it degenerated} quantum system., i.e., to send quantum information
without sending a clock.
This description may  be a good approximation for the more realistic
physical situation that the systems that Alice can send to Bob
evolve on a time scale that is considerably slower  than the 
time scale of the considered synchronization.

The following protocol allows Bob to obtain the clock $(\rho,1\otimes \beta)$
if one tells Alice the absolute time.

\begin{enumerate}

\item Tell Alice that the true time is $t_0 >0$.

\item Alice implements the transformation $\alpha_{-t_0}$ on her system
and puts her part of the joint state into a physical system
that is equal to her original one with the only difference that its 
time evolution
is trivial. In the classical situation, she uses a system with the
 same phase space as the original, in the quantum case an isomorphic
Hilbert space with the Hamiltonian $H=0$ can be used.
She sends this `frozen clock' to Bob. This has not necessarily to be 
performed by transferring a real physical system. In the quantum case
Alice can teleport this clock using prior entanglement on degenerated systems.
In the classical case she can phone Bob and tell him the state
of her frozen clock.

\end{enumerate}

If the initial joint state is $\rho$ (at the time $0$) 
then Bob will obtain the state
\[
(1\otimes \beta_{t-t_0})\circ (\alpha_{-t_0} \otimes 1) \circ 
(\alpha_{t_0} \otimes \beta_{t_0}) (\rho)=
(1 \otimes\beta_t) (\rho) 
\]
at the time $t$. 
Formally he has the clock $(\rho,1\otimes \beta)$.
To see that this is the best clock that Bob can obtain we show
first that each clock defines a synchronism as follows.
Assume that  Bob has an arbitrary clock $(\sigma,\gamma)$ with period 
$1$, i.e., at the time $t$ the state is $\gamma_t (\sigma)$.
Alice can read the true time $t\in [0,1)$ on her clock without any
error.
Now we consider an observer who does not know anything
about the true time. From his point of view the state of the joint system
is stationary and given by an equal distribution over
all states of the form $\exp(i2\pi t) \otimes \gamma_t(\rho)$.
This is a non-trivial synchronism of the bipartite system.
If Bob starts with the clock $(\sigma,\gamma)=(\rho,1\otimes \beta)$
we can obtain the synchronism $(\rho,\alpha,\beta)$ by the following 
protocol (obtained by reversing the scheme above).  
(1) Bob  sends
 the left part of the state $\rho$ on a stationary system to Bob.
(2) Alice puts the left part of the state $\rho$ on a system
with dynamical evolution $\alpha$ and applies an additional transformation
$\alpha_{t_0}$  to the system according to the present value $t_0$ of her
absolute clock. Although we omit the formal proof this protocol shows
that the following two situations are equivalent 
with respect to their synchronism:

\begin{enumerate}

\item
Bob has the clock  $(\rho, 1\otimes \beta)$ and Alice has a perfect classical 
clock. 

\item
The state  $\rho$ is the joint state of Alice and Bob with the time
evolution $\alpha$ and $\beta$ on Alice's and Bob's system,
 respectively.

\end{enumerate}

These considerations define a one-to-one correspondence between 
equivalence classes of synchronism and equivalence classes of clocks.
The fact that the clock obtained from a synchronism
can be used to recover the original synchronism shows that
$(\rho,1\otimes \beta)$ is indeed the best achievable clock 
from the synchronism $(\rho,\alpha,\beta)$. Otherwise
the better clock $(\sigma,\gamma)$ would allow to obtain a better
synchronism than the resource $(\rho,\alpha,\beta)$.

The difference between the clock that Bob can obtain if we tell Alice
the true time and the clock Alice can obtain if we tell Bob the time
can easily be understood by comparing
$(\rho,\alpha^{-1/2} \otimes \beta^{1/2})$ to
$(\rho,\alpha^{1/2} \otimes \beta^{-1/2})$. Formally, 
Alice's clock is the time-reversed of Bob's clock.
Consider the classical unit circle clocks above. Then
Bob obtains the probability density $q$ whereas Alice
would obtain the probability density $z \mapsto q(z^{-1})$.
The clock defined by such a distribution is equivalent to
a clock with probability density $q$ with time-reversed dynamical
evolution.

The arguments above show that the problem of evaluating synchronizations
can completely be reduced to the problem of evaluating clocks. 
One could for instance use Fisher timing information to describe
the accuracy of synchronism.

\section{Conclusions}

The main idea of the quasi-order of clocks is to consider 
coherent superpositions of energy eigenstates as a resource carrying
information about the time at which the state has been prepared.
The source of timing information can also be a sender transmitting a 
signal, for instance a time-localized pulse. 
The quasi-order of clocks describes the fact that nobody can change 
the signal  in such a way that the information about the sender's time is 
increased except he receives additional hints about the sender's time.
This implies lower bounds on the achievable fidelity if a two-level system 
should be driven in a
superposition state by a coherent laser field.

Furthermore the quasi-order gives restrictions to the possibility
of copying timing information:
If a device produces two outgoing signals from an input signal, the timing 
information of the outgoing signals cannot be larger than the original one.
This does not mean necessarily that the timing information of each of both
outgoing signals is less than the original one. The outgoing  pulses 
might be perfectly correlated in time such that the
timing information of each component coincides with the information 
inherent in both. However, perfect correlations are only
possible in the limit of infinite signal energy.
Otherwise at least one of the signals will be less accurately localized in 
time. We quantified the time accuracy of  signals by 
the statistical distance of the actual state and the state after
an infinitesimal time and proved a lower bound on the
loss  of accuracy in the signal cloning process.  
Since this lower bound depends on the total energy of the outgoing signals 
we have found a constraint for low-power signal  processing.

Taking a thermodynamic point of view, one might state that
coherent superpositions of energy states are a resource 
which have to be supplied by a source of timing information.

In the same way as timing information is a resource
for a single party  system the {\it relative} timing information
or {\it synchronism} of a bipartite system is a resource 
if no exchange of clocks
is allowed. The problem of classifying bipartite systems with respect
to their synchronism is shown to be equivalent to the problem
of classifying clocks provided that it is allowed
to exchange {\it stationary} quantum or classical systems.

\end{multicols}

\section*{Acknowledgments}

Thanks to J\"{o}rn M\"{u}ller-Quade for useful remarks.
This work has been supported by DFG-grants of the project
`Komplexit\"{a}t und Energie'.


\begin{thebibliography}{10}

\bibitem{GLMS}
V.~Giovannetti, S.~Lloyd, L.~Maccone, and M.~Shahriar.
\newblock Physical limits to clock synchronization.
\newblock {\em LANL preprint},
\newblock quant-ph/0110156.

\bibitem{FMWW}
H.~Fang, K.~Matsumoto, X.~Wang, and M.~Wadati.
\newblock Quantum cloning machnines for equitorial qubits.
\newblock {\em LANL preprint}, (quant-ph/0101101).

\bibitem{Fu96}
C.~Fuchs.
\newblock Distinguishability and accessible information in quantum theory.
\newblock {\em LANL preprint}
\newblock (quant-ph/9601020), 1995.

\bibitem{JWZ00}
D.~Janzing, P.~Wocjan, R.~Zeier, R.~Geiss, and T.~Beth.
\newblock Thermodynamic cost of reliability and low temperatures: Tightening
  landauer's principle and the second law.
\newblock {\em Int. Jour. Theor. Phys.}, 39(12):2217--2753, 2000.
\newblock {\em LANL preprint} quant-ph/0002048.

\bibitem{BR1}
O.~Bratteli and D.~Robinson.
\newblock {\em Operator algebras and quantum statistical mechanics}.
\newblock Springer, New York, 1987.

\bibitem{PS}
A.~Peres and P.~Scudo.
\newblock Covariant quantum measures may not be optimal.
\newblock {\em LANL preprint}
\newblock quant-ph/0107114.

\bibitem{Da78}
E.~Davies.
\newblock Information and quantum measurement.
\newblock {\em IEEE Trans. Inform. Theory}, IT-24(5):596--599, 1978.

\bibitem{SBJOH}
M.~Sasaki, S.~Barnett, R.~Josza, M.~Osaki, and O.~Hirota.
\newblock Accessible information and optimal strategies for real symmetrical
  quantum sources.
\newblock {\em LANL-preprint}, quant-ph/9812062v3).

\bibitem{WM}
D.~Walls and G.~Milburn.
\newblock {\em Quantum optics}.
\newblock Springer, Heidelberg, 1994.

\bibitem{ML98}
N.~Margolus and L.~Levitin.
\newblock The maximum speed of dynamical evolution.
\newblock {\em LANL preprint} quant-ph/9710043 v2.

\bibitem{BC94}
S.~Braunstein and C.~Caves.
\newblock Statistical distance and the geometry of quantum states.
\newblock {\em Phys. Rev Lett.}, 72(22):3439, 1994.

\bibitem{BCM96}
B.~Braunstein, C.~Caves, and G.~Milburn.
\newblock Generalized uncertainty relations: Theory, examples and lorentz
  invariance.
\newblock {\em Ann. Phys.}, 247:135--173, 1996.

\bibitem{Hole}
A.~Holevo.
\newblock {\em Probabilistic and Statistical Aspects of Quantum Theory}.
\newblock North-Holland, Amsterdam, 1982.

\bibitem{Hel}
C.~Helstrom.
\newblock {\em Quantum Detection and Estimation Theory}.
\newblock Academic Press, New York, 1976.

\bibitem{Cra}
H.~Cram\'{e}r.
\newblock {\em Mathematical methods of statistics}.
\newblock Princeton University, Princeton, 1946.

\bibitem{GVV89}
D.~Goderis, A.~Verbeure, and P.~Vets.
\newblock Non-commutative central limits.
\newblock {\em Prob. Th. Rel. Fields}, 1989.

\bibitem{GV89}
D.~Goderis and P.~Vets.
\newblock Central limit theorem for mixing quantum systems.
\newblock {\em Comm. Math. Phys.}, 1989.

\bibitem{Ho73}
A.~Holevo.
\newblock Statistical decision theory for quantum systems.
\newblock {\em J. Multivar. Analys.}, 1973.

\end{thebibliography}
\end{document}